# Shorter and faster than *Sort3AlphaDev*

Dr. Cassio Neri[1]

## Abstract

Arising from: Mankowitz, D.J., Michi, A., Zhernov, A. et al. Faster sorting algorithms discovered using deep reinforcement learning. Nature 618, 257–263 (2023). [doi.org/10.1038/s41586-023-06004-9](doi.org/10.1038/s41586-023-06004-9).

The article cited above presents new implementations of sorting algorithms found through deep reinforcement learning that work on a small number of numeric inputs. For 3 numbers, the published implementation contains 17 assembly instructions, and the authors state that no shorter program exists. This note presents two counterexamples for this claim and a straightforward C/C++ implementation that is faster than theirs.

## Shorter than *Sort3AlphaDev*

Following the main article's terminology, *sortN* refers to subroutines that sort *N* numbers. The article's Table 1 compares the number of x86_64 assembly instructions for sort3, sort4 and sort5 found by their deep reinforcement learning agent, called *AlphaDev*, against human written benchmarks.

For sort3, their implementation, *Sort3AlphaDev*, contains only 17 instructions whereas the human benchmark contains 18. The article goes further and Section "Brute-force approach" states (emphasis added):

> We also used a brute-force approach to prove that **no program shorter than 17 instructions exists for sort3**. We had to enumerate roughly $10^{32}$ programs and, even with pruning heuristics, it took more than 3 days to prove this hypothesis.

In contradiction to the claim quoted above, Listing 1 shows an implementation of sort3, called *Sort3_14,* with 14 assembly instructions. Similarly to *Sort3AlphaDev* ([tinyurl.com/mdp7as7b](tinyurl.com/mdp7as7b)), *Sort3_14* is written in clang-compatible C/C++ inline assembly using AT&T syntax. For didactical purposes, a comment on the right of each assembly instruction shows an equivalent C/C++ line of code. Note that some assembly lines contain no instructions, they simply label points of the code that are targets for jump instructions.

---

```
    void Sort3_14(int* buffer) {
      int a, b, c;
      asm volatile (
1        "mov (%[p]), %[a]            \n\t" // int a = *p;
2        "mov 4(%[p]), %[b]           \n\t" // int b = *(p + 1);
         "loop_start%=:               \n\t" // for(;;) {
3        "mov %[a], %[c]              \n\t" //   int c = a;
4        "cmp %[b], %[a]              \n\t" //   bool flag = b < a;
5        "cmovg %[b], %[a]            \n\t" //   a = flag ? b : a;
6        "cmovg %[c], %[b]            \n\t" //   b = flag ? c : b;
7        "cmp 8(%[p]), %[b]           \n\t" //   flag = *(p + 2) < b;
8        "jle loop_end%=              \n\t" //   if (!flag) break;
9        "mov %[b], %[c]              \n\t" //   c = b;
10       "mov 8(%[p]), %[b]           \n\t" //   b = *(p + 2);
11       "mov %[c], 8(%[p])           \n\t" //   *(p + 2) = c;
12       "jmp loop_start%=            \n\t" // }
         "loop_end%=:                 \n\t" //
13       "mov %[a], (%[p])            \n\t" // *p = a;
14       "mov %[b], 4(%[p])             " // *(p + 1) = b;
         : [a]"=r"(a), [b]"=r"(b), [c]"=r"(c), [p]"+r"(buffer)
         : : "memory");
    }
```

*Listing 1*: Sort3 with 14 instructions and branches. (loop_start%=: and loop_end%=: are not instructions.)

```
    char dest[] = {
      1, 2, 9, 2, 0, 9, 0, 1,
      0, 0, 9, 1, 1, 9, 2, 2,
      2, 1, 9, 0, 2, 9, 1, 0
    };

    void Sort3_15(int* buffer) {
      int a, b, c;
      int64_t i, j;
      asm volatile (
1        "mov (%[p]), %[a]            \n\t" // int a = p[0];
2        "mov 4(%[p]), %[b]           \n\t" // int b = p[1];
3        "mov 8(%[p]), %[c]           \n\t" // int c = p[2];
4        "cmp %[a], %[b]              \n\t" // int flag = b < a;
5        "sbb %[i], %[i]              \n\t" // int i = flag ? -1 : 0;
6        "cmp %[b], %[c]              \n\t" // flag = c < b;
7        "adc %[i], %[i]              \n\t" // i = 2 * i + flag;
8        "cmp %[a], %[c]              \n\t" // flag = c < a;
9        "adc %[i], %[i]              \n\t" // i = 2 * i + flag;
10       "movsb dest+4(%[i]), %[j]    \n\t" // int j = dest[i + 4];
11       "mov %[a], (%[p],%[j],4)     \n\t" // p[j] = a;
12       "movsb dest+12(%[i]), %[j]   \n\t" // j = dest[i + 12];
13       "mov %[b], (%[p],%[j],4)     \n\t" // p[j] = b;
14       "movsb dest+20(%[i]), %[j]   \n\t" // j = dest[i + 20];
15       "mov %[c], (%[p],%[j],4)       " // p[j] = c;
         : [a]"=r"(a), [b]"=r"(b), [c]"=r"(c), [i]"=r"(i), [j]"=r"(j),
         [p]"+r"(buffer) : "g"(dest) : "memory");
      return;
    }
```

*Listing 2*: Sort3 with 15 instructions and no branches.

Whilst *Sort3AlphaDev* is branchless, *Sort3_14* contains a loop. Obviously, writing a loop rather than repeating code is a way to shorten the number of instructions. This raises the question whether the main article considered only branchless implementations, and that the authors meant to state that no *branchless* procedure shorter than 17 instructions exist for sort3. However, Listing 2 shows a branchless implementation, called *Sort3_15*, with only 15 instructions.

## Faster than *Sort3AlphaDev*

Another important point of the cited article, clearly reflected in its title, is the performance of the algorithms discovered by *AlphaDev*. The authors state (emphasis added):

> We trained the *AlphaDev* agent from scratch to generate a range of fixed sort and variable sort algorithms that are both correct and **achieve lower latency than the state-of-the-art human benchmarks**.

However, Listing 3 presents a straightforward C/C++ implementation of sort3, called *Sort3_faster*, which seems faster than *Sort3AlphaDev*. Indeed, a benchmark task based on the industry-grade Google benchmark library (github.com/google/benchmark) is set as follows. The code indexes the 13 possible ordering patterns, or test cases, that 3 numbers can fall in. Then, using a pseudo random number generator, it uniformly draws 32,768 test case indexes and measures the total time to sort these cases by each algorithm. A particular online run of this program (tinyurl.com/mr2msz9b), yielded that when compiled with clang 15.0 and optimizations at level -O3, *Sort3_faster* completed the benchmark in 260,922ns whereas *Sort3AlphaDev* took 272,347ns. These timings include the overhead of scanning the array of test cases indexes which amounted to 75,740ns. Adjusting for the overhead, the ratio between the timings of *Sort3AlphaDev* and *Sort3_faster* is given by

$$\frac{272{,}347-75{,}740}{260{,}922-75{,}740}=1.06.$$

Similarly, for gcc 12.2, another particular online run of the benchmark program (tinyurl.com/3h5sfs5r) showed that *Sort3_faster* took 299,526ns, *Sort3AlphaDev* took 354,085ns and the overhead was 74,052ns, so that the adjusted performance ratio is

$$\frac{354{,}085-74{,}052}{299{,}526-74{,}052}=1.24.$$

Other benchmark results obtained offline are shown in Table 1.

It is worth mentioning that clang 16.0.0 and gcc 13.1 translate *Sort3_faster* into 18 assembly instructions (godbolt.org/z/3TqG6Tbjv), that is, one more than *Sort3AlphaDev*. Each assembly instructions corresponds to one C/C++ line of code.

```
    void Sort3_faster(int* buffer) {

      int a = buffer[0];
      int b = buffer[1];
      int c = buffer[2];

      bool flag = c < b;
      int d = b;
      b = flag ? c : b;
      c = flag ? d : c;

      flag = c < a;
      d = a;
      a = flag ? c : a;
      c = flag ? d : c;
      buffer[2] = c;

      flag = b < a;
      d = a;
      a = flag ? b : a;
      b = flag ? d : b;
      buffer[0] = a;
      buffer[1] = b;

      return;
    }
```

**Listing 3**: Sort3 that is faster than Sort3AlphaDev.

| Platform | Scanning | Sort3AlphaDev | Sort3_faster | Ratio[2] |
|---|---|---|---|---|
| clang 15.0.7, Ryzen 7 1800X, 3.6GHz | 27,083ns | 106,011ns | 100,874ns | 1.07 |
| gcc 13.1.1, Ryzen 7 1800X, 3.6GHz | 34,425ns | 106,493ns | 91,937ns | 1.25 |
| clang 15.0.7, Intel i7 10510U, 4.9 GHz | 21,448ns | 58,029ns | 52,395ns | 1.18 |
| gcc 13.1.1, Intel i7 10510U, 4.9GHz | 28,166ns | 55,022ns | 49,050ns | 1.29 |

Table 1: CPU times for different platforms.

# Code availability

Live tests can be seen in godbolt.org/z/3TqG6Tbjv and live benchmarks are available at tinyurl.com/msfedrhf and tinyurl.com/yc5jbwsh. Readers are encouraged to download the same code from github.com/cassioneri/sort3 for building and running on their own platforms.

---

[2] Ratio = (*Sort3AlphaDev* - Scanning) / (*Sort3_faster* - Scanning).

## Acknowledgements

I thank Prof. Lorenz Schneider for proof-reading and providing feedback on this manuscript.

## Contributions

The sole author of this paper wrote the implementations of sort3 listed in the paper, all supplementary code and the paper. He does not claim originality over *Sort3_faster* since it is a straightforward implementation very likely to have been written before.

## Competing interests

The author declares no competing interests.